\documentclass[12pt,preprint]{aastex}
\usepackage{epsfig}
\usepackage{graphicx}
\usepackage{float}
\usepackage{amsmath}
\usepackage{color}
\usepackage{amssymb}
\usepackage{amsfonts}
\usepackage{units}
\usepackage[colorlinks,linkcolor=blue,anchorcolor=green,citecolor=blue]{hyperref}

\shorttitle{Polarization properties of magnetar giant flare pulsating tails}
\shortauthors{Yang \& Zhang}

\begin{document}

\title{On the polarization properties of magnetar giant flare pulsating tails}
\author{Yuan-Pei Yang\altaffilmark{1,2,3} and Bing Zhang\altaffilmark{3,4,5}}

\affil{$^1$School of Astronomy and Space Science, Nanjing University, Nanjing 210093, China; yypspore@gmail.com;\\
$^2$ Key Laboratory of Modern Astronomy and Astrophysics (Nanjing University), Ministry of Education, China; \\
$^3$ Department of Physics and Astronomy, University of Nevada, Las Vegas, NV 89154, USA; zhang@physics.unlv.edu; \\
$^4$ Department of Astronomy, School of Physics, Peking University, Beijing 100871, China; \\
$^5$ Kavli Institute of Astronomy and Astrophysics, Peking University, Beijing 100871, China}

\begin{abstract}
Three giant flares have been detected so far from soft gamma-ray repeaters, each characterized by an initial short hard spike and a pulsating tail. The observed pulsating tails are characterized by a duration of $\sim100\,\unit{s}$, an isotropic energy of $\sim 10^{44}\,\unit{erg}$, and a pulse period of a few seconds. The pulsating tail emission likely originates from the residual energy after the intense energy release during the initial spike, which forms a trapped fireball composed of a photon-pair plasma in a closed field line region of the magnetars. Observationally the spectra of pulsating tails can be fitted by the superposition of a thermal component and a power-law component, with the thermal component dominating the emission in the early and late stages of the pulsating tail observations. In this paper, assuming that the trapped fireball is from a closed field line region in the magnetosphere, we calculate the atmosphere structure of the optically-thick trapped fireball and the polarization properties of the trapped fireball.
By properly treating the photon propagation in a hot, highly magnetized, electron-positron pair plasma, we tally photons in two modes (O mode and E mode) at a certain observational angle through Monte Carlo simulations. Our results suggest that the polarization degree depends on the viewing angle with respect to the magnetic axis of the magnetar, and can be as high as $\Pi\simeq30\%$ in the $1-30\,\unit{keV}$ band, and $\Pi\simeq10\%$ in the $30-100\,\unit{keV}$ band, if the line of sight is perpendicular to the magnetic axis. 
\end{abstract}


\section{Introduction}

Magnetars are neutron stars with super-strong magnetic fields. The typical magnetic strength of magnetars is of the order of $(10^{14}-10^{15}) \unit{G}$, which is higher than the critical magnetic field $B_{Q}=m_e^{2}c^{3}/\hbar e\approx 4.4\times 10^{13}\unit{G}$, and is $\sim (10^{2}-10^{3})$ times stronger than that of normal pulsars. Historically, they were discovered in two manifestations, i.e. anomalous X-ray pulsars (AXPs) and soft gamma-ray repeaters (SGRs), even though later observations suggested that there seems no clear distinction between the two sub-categories. SGRs are discovered in the hard X-ray/soft gamma-ray band characterized by repeating bursts. Eleven SGRs have been confirmed so far \citep{ola14}. Three of them each displayed one giant flare (SGR 0526-66 on March 5, 1979; SGR 1900+14 on August 27, 1998; SGR 1806-20 on December 27, 2004). These giant flares are characterized by an initial short-hard spike followed by a pulsating tail \citep{mer08}. The initial spike has a duration of $\sim 0.1\unit{s}$, and an isotropic energy of $(10^{44}-10^{46}) \unit{erg}$. The pulsating tail, on the other hand, shows a decay pattern and is characterized by a typical duration of $\sim 100\,\unit{s}$, an isotropic energy of $\sim 10^{44}\,\unit{erg}$, and a pulsating period of a few seconds (the magnetar spin period). The spectrum of the early spike is characterized by a temperature of $T\sim100\,\unit{keV}$, which is much harder than that of normal SGR bursts. The spectrum of the pulsating tail is softer than that of the initial spike, which may be fitted with the superposition of a thermal component with $T\sim10-30\,\unit{keV}$ and a non-thermal power law component. The decay light curves of the pulsating tail typically show time-evolving, complex pulse profiles.

As pointed out by \citet{tho95,tho01}, the tail emission likely originates from the residual energy of the explosion that gives rise to the initial spike. Owing to the large energy density and strong magnetic field pressure in the magnetar magnetosphere, the residual energy is confined in the magnetosphere in the form of an optically thick photon-pair plasma named a ``trapped fireball''. Such a trapped fireball would be the source of the thermal emission component observed in the pulsating tail. The power law spectral component may be formed from the jets associated with the open-field lines that are outside the trapped fireballs \citep{tho01}. The structure of the trapped fireball may determine the pulse profile of the pulsating tail. In fact, the magnetars show more complicated pulse structure profiles during the giant flare decaying phase than during the quiescent phase, the latter of which typically shows a single, broad-peak, nearly sinusoidal pulse shape. For example, the decay phase of the giant flare of SGR 0526-66 on March 5, 1979 showed a double-peak pulse profile \citep{maz82}; the giant flare of SGR 1900+14 on August 27, 1998 showed a complex four-peak profile \citep{hur99,fer01}; and the giant flare of SGR 1806-20 on December 27, 2004 showed a roughly three-peak profile \citep{hur05,bog07}. On the other hand, the temperature of the phase-resolved spectra during the decay phase shows a different evolution pattern from the light curve: For SGR 1900+14, the phase-resolved temperature shows two peaks in each period, with the epoch of the highest thermal flux corresponding to the lowest count rate in the light curve \citep{fer99}; For SGR 1806-20, on the other hand, the evolution of temperature seems to track the photon count in the light curve \citep{bog07}. Therefore, the true structure of the trapped fireball is likely very complex. The effective blackbody radius inferred from the data suggests that the radius of the trapped fireball is a few times of the magnetar radius \citep{bog07}, which suggests that the trapped fireball should be in the magnetosphere rather than being a hot spot on the magnetar surface. A natural idea is that the trapped fireball is a closed-field region in the magnetosphere (and the non-thermal emission comes from the open field line regions). The multi-peak profile of the light curves might be the result of the inhomogeneous distribution of the plasmas within the trapped fireball and the non-thermal jets within the magnetosphere. Due to the complexity of the problem itself, it is not easy to fully understand the mechanism of the giant flares. Nonetheless, a simple model that assumes that the trapped fireball is from a closed field region in the magnetosphere may catch the essence of the problem. 

Polarization is a key element to test theoretical models of the physical processes and geometric structure of magnetars. Unfortunately, none of the three giant flares were detected with polarimeters, so that their polarization properties remain a mystery. Thanks to technological development and funding support, a few X-ray/$\gamma$-ray polarimeters are imminent, which would make ground-breaking discoveries to study the underlying physics in many astrophysical objects, such as isolated neutron stars, magnetars, X-ray pulsars, pulsar wind nebulae, rotation-powered pulsars, and millisecond X-ray pulsars \citep{fer11b}. For example, in the conventional X-ray band ($2-10\,\unit{keV}$), 
polarimeters that are under construction include Imaging X-ray Polarimetry Explorer (IXPE, \citet{wei14}), Polarimeter for Relativistic Astrophysical X-raY Sources (PRAXyS, \citet{jah15}). 
For hard X-rays and soft-$\gamma$-rays (above $10\,\unit{keV}$), several detector concepts have been proposed. For example, there has been an extensive study of the POET (Polarimeters for Energetic Transients) mission concept \citep{mcc14} as well as its extension LEAP (LargE Area Polarimeter). On the other hand, the POLAR detector \citep{pro05,xia15}, a Compton polarimeter designed to measure linear polarization of gamma-rays from gamma ray bursts (GRBs), is scheduled to be launched with the Chinese Space Station Tiangong-2 in 2016. 
POLAR has a large effective area of $400\,\unit{cm^2}$ and a field-of-view covering about one third of the sky. Its energy band covers from $50\,\unit{keV}$ to $500\,\unit{keV}$, which can cover part of the energy range of the pulsating tail of SGR giant flares. For a total flux larger than $10^{-5}\,\unit{erg\,cm^{-2}}$, POLAR can perform a measure of polarization degree with a relative systematic error $\sim10\%$ \citep{xio09,sua10}. It would be an ideal detector for giant flare polarization studies, should such a giant flare occurs during its mission period.

Among the available detectors, the Reuven Ramaty High Energy Solar Spectroscopic Imager (RHESSI) is an instrument designed to study the brightest solar flares, which has the polarization detecting capability  \citep{mcc07}. Interestingly, the giant flare of SGR 1806-20 on December 27, 2004 was detected with RHESSI \citep{bog07}. However, no polarization study of the pulsating tail emission of this event is published.

Many authors have calculated the polarization properties from neutron stars. \citet{pav00} calculated the polarization of thermal X-rays from isolated neutron stars with strong magnetic fields. \citet{ho01,ho03} studied radiation spectra from the atmosphere of neutron stars and the effect of vacuum polarization. \citet{hey02} and \citet{hey03} suggested that the birefringence of the magnetized QED vacuum decouples the polarization modes up to a large radius, so that the surface polarization of neutron stars can be sufficiently large to be observed. \citet{ade06} and \citet{ade09} calculated the soft X-ray polarization in thermal magnetar emission from a hot spot centered around the magnetic dipole axis. \citet{fer11} studied the X-ray polarization signature of the magnetosphere of quiescent magnetars. \citet{tav14,tav15} studied how future X-ray polarization measurements by the next generation detectors may be used to probe the magnetar magnetospheric structure. 

In this paper, we study the atmospheric structure and polarization properties of trapped fireballs during the decay phase of giant flares. Since the radiation mechanism of the non-thermal component is still uncertain, we mainly consider the thermal emission component from the trapped fireball. Observationally, the thermal radiation component seems to dominate the observed emission during most phases in the decaying tail, and the flux level of the thermal component is higher than that of the power-law component between $10\,\unit{keV}$ and $100\,\unit{keV}$ \citep{bog07}. 
The paper is organized as follows. In Sect. 2, we introduce a simple model of the atmospheric structure of the trapped fireball. In Sect. 3, the photon polarization properties are calculated by considering dielectric tensor of the electron-positron pair plasma as well as the effect of vacuum polarization, and the scattering opacities for the two photon modes are presented. In Sect. 4. we calculate the polarization degree of the trapped fireball using Monte Carlo simulations. The results are summarized in Sect. 5 with some discussion.

\section{Atmospheric Structure of Trapped Fireball}

In this section, we calculate the atmospheric structure of the trapped fireball in a simplified model. We assume a dipolar configuration for the magnetar magnetic fields, and take the region enclosed by a closed field line (not necessarily the last closed field line) as the trapped fireball (Fig.1). Thanks to the extreme large optical depth of the trapped fireball, we only need to solve the equations in a relatively thin layer (the bisque layer shown in Fig.1), which corresponds to the radius range from an optical depth $\tau \gg 1$ (bottom of the atmosphere of the trapped fireball) to $\tau \ll 1$. The observed photons mostly come from this region. The structure of this thin atmosphere can be solved via a set of equations, including the magnetic dipole field equation, the equations of state, the hydrostatic equilibrium equation, the energy transport equation, and the optical depth equation.

The total energy radiated in the pulsating tail of a giant flare is $E_{tail}\sim10^{44}\unit{erg}$ \citep{mer08}. The requirement for the energy being trapped by magnetic fields demands
\begin{eqnarray}
\frac{B_{R_s+\Delta R}^2}{8\pi}\gtrsim\frac{E_{tail}}{3\Delta R^3},\label{trap}
\end{eqnarray}
where $R_s$ is the radius of the neutron star, and $\Delta R$ is the scale of the trapped fireball (Fig.1). The factor $1/3$ comes from converting the internal energy density (both the thermal energy\footnote{Even though at the photosphere the electron temperature is 10s of keV, in the bulk of the trapped fireball, the electron temperature is higher than 511 keV, so that the fireball can be regarded as a relativistic gas. See detailed discussion below.} and the radiation energy) of the trapped fireball to pressure. For a dipolar field, one has $B(r)=B_\ast (r/R_s)^{-3}$, where $B_\ast$ is the characteristic surface magnetic field strength of the neutron star. Observations show that the effective radius of thermal emission is a few times of the magnetar radius, $\Delta R\gtrsim R_s$ \citep{bog07}. We therefore require that the magnetic field strength satisfies
\begin{eqnarray}
B_\ast\gtrsim 2\times10^{14}\unit{G}\left(\frac{\Delta R}{10\unit{km}}\right)^{-3/2}\left(\frac{E_{tail}}{10^{44}\unit{erg}}\right)^{1/2}\left(\frac{1+\Delta R/R_s}{2}\right)^3 > B_Q,
\end{eqnarray}
which suggests that the neutron star is a magnetar. 
In the closed field line region, the electron-positron pair plasma is trapped, since transport of charged particles across field lines is suppressed by strong magnetic fields, and since gravity is no longer the dominant confining force. The critical trapping luminosity $L_c$ can be estimated by the balance between the radiative energy density and the magnetic energy density $L/4\pi r^2 c\sim B^2/8\pi$, i.e.
\begin{eqnarray}
L_c\simeq 3\times10^{49}\,\unit{erg\,s^{-1}}\left(\frac{B}{B_Q}\right)^2\left(\frac{r}{R_s}\right)^2,
\end{eqnarray}
which is much larger than the luminosity of the giant flares.

For a magnetic dipole field, the strength of the magnetic field at a point $(r,\theta)$ is
\begin{eqnarray}
B(r,\theta)=\frac{B_p}{2}\left(\frac{R_s}{r}\right)^3\sqrt{1+3\cos^2(\theta)},\label{dipole}
\end{eqnarray} 
where $B_p$ is the polar cap magnetic field strength at the surface. The border of the closed magnetosphere field lines that enclose the trapped fireball is given by $r=R_{max}\sin^2\theta$, where $R_{max}$ is the maximum radius. 

Next, we consider the equation of state in such a trapped fireball. The deposition of $E_{tail}\sim10^{44}\,\unit{erg}$ in the magnetosphere is sufficient to generate a hot photon-pair plasma. For the internal region of the trapped fireball, the electron-positron pairs are mildly-relativistic. The total energy density of the photon-pair plasma is $U=U_\gamma+U_{e^\pm}=(11/4)aT^4$ (see Eq. (\ref{a8}) in Appendix). The temperature in the internal region of the fireball is $T=1.3\,\unit{MeV}(E_{tail}/10^{44}\,\unit{erg})^{1/4} (\Delta R/R_s)^{-3/4}$. For the atmosphere of the trapped fireball, on the other hand, observations show that its effective temperature is $k_BT_{eff}\sim30\,\unit{keV}\ll m_ec^2$, which is much less than the first Landau level energy $\hbar eB/m_ec=508\,\unit{keV}(B/B_Q)$. Therefore, this is a one-dimensional, magnetized, and non-relativistic electron-positron pair plasma, whose density is given by (see Eq. (\ref{a9}) in Appendix)
\begin{eqnarray}
\rho=m_e\frac{(m_ec)^3}{\hbar^3(2\pi^3)^{1/2}}\frac{B}{B_Q}\left(\frac{k_BT}{m_ec^2}\right)^{1/2}\exp(-m_ec^2/k_BT).\label{eos}
\end{eqnarray}
We apply an ideal gas equation of state and assume that the gas is in local thermodynamic equilibrium with the radiation field. The total pressure $P$ is given by the sum of the gas pressure $P_g$, radiation pressure $P_\gamma$, and magnetic pressure $P_B$, i.e.
\begin{eqnarray}
P=P_g+P_r+P_B,\,\,\,P_g=\frac{\rho}{m_e}k_BT,\,\,\,P_r=\frac{4\sigma_{SB}}{3c}T^4,\,\,\,P_B=\frac{B^2}{8\pi}. \label{pressure}
\end{eqnarray}
The giant flare pulsating tails are characterized by a duration of $\sim100\,\unit{s}$. We assume that the trapped fireball is in hydrostatic equilibrium during the pulsating tail phase:
\begin{eqnarray}
\nabla P+\rho \nabla\Phi-\frac{1}{4\pi}(B\cdot\nabla)B=0,
\end{eqnarray}
where $\Phi$ is the gravitational potential. For the dipolar magnetic field, the magnetic stress in radial direction ($r$ direction) is given by
\begin{eqnarray}
F_B\equiv\frac{1}{4\pi}[(B\cdot\nabla)B]_{r}=-\frac{1}{2\pi}\left(\frac{B_p}{2}\right)^2\frac{R_s^6}{r^7}[1+5\cos^2(\theta)].\label{stress}
\end{eqnarray}
Therefore one has
\begin{eqnarray}
\frac{dP}{dr}=-\frac{GM_s\rho}{r^2}\left(1-\frac{R_g}{r}\right)^{-1}\left[1+\frac{P+1.5P_g+3P_r+P_B}{\rho c^2}\right]+F_B,
\end{eqnarray}
where $M_s$ is the neutron star mass (with the trapped fireball mass neglected), and $R_g=2GM_s/c^2$ is the gravitational radius. Here, we neglect the term $4\pi r^3P/M_sc^2$ in the general Oppenheimer-Volkoff equation, because its contribution is always smaller than $10^{-4}$ in our atmosphere model. In order to solve the atmospheric structure, we also need the temperature gradient equation
\begin{eqnarray}
\frac{dT}{dr}=\frac{T}{P}\frac{dP}{dr}\nabla,
\end{eqnarray}
where $\nabla\equiv(\ln T/\ln P)$ is determined by the energy transport equation. As pointed out by \citet{raj96}, convection is strongly suppressed by strong magnetic fields of magnetars. We therefore only consider radiative transport in this paper.  
For simplicity of treatment and without losing generality, we use the result from the radiative transport in spherical symmetry \citep{tho77,pac86}, which is given by
\begin{eqnarray}
\nabla\equiv\left(\frac{\ln T}{\ln P}\right)_r&=&\left[\frac{\kappa L}{16\pi cGM_s}\left(\frac{P}{P_r}\right)\left(1-\frac{R_g}{r}\right)^{-1/2}+\frac{P}{\rho c^2}\right]\nonumber\\
&\times&\left[1+\frac{P+1.5P_g+3P_r+P_B}{\rho c^2}\right]^{-1},
\end{eqnarray}
where $L$ is the luminosity of the photosphere (i.e. the observed luminosity of the pulsating tail), and $\kappa$ is the Rosseland mean opacity, which is discussed in the next section. In order to obtain the scale length of the photosphere, one also needs the optical depth equation
\begin{eqnarray}
\frac{d\tau}{dr}=-\kappa\rho \left(1-\frac{R_g}{r} \right)^{-1/2}.\label{optical}
\end{eqnarray}
Finally, we introduce the boundary of the bottom of the atmosphere that is defined by $(R_{max,0},T_0)$, where $R_{max,0}$ is the maximum radius of the closed line of the atmosphere bottom, and $T_0$ is the temperature at the bottom of the atmosphere. Using Eqs.(\ref{dipole})-(\ref{pressure}) and Eqs.(\ref{stress})-(\ref{optical}), we can then solve for the atmospheric structure of the trapped fireball.

We assume $M_s=1.44\,M_{\odot}$, $R_s=10^6\,\unit{cm}$, $B_p=10^{15}\,\unit{G}$, the rotation period of the magnetar $P=5\,\unit{s}$, and the luminosity of the pulsating tail $L=10^{42}\,\unit{erg\,s^{-1}}$. Since observations show that the effective radius of the thermal emission is a few times of the radius of the magnetar, we assume that the maximum radius of the bottom of the atmosphere is $R_{max,0}=2R_s$. The corresponding temperature $T_0$ at $R_{max,0}$ should be less than the pair production temperature $T_\pm\simeq 6\times10^9\,\unit{K}$ in order to reach a quasi-steady state solution.
In our model, we take $T_{0}=10^9\,\unit{K}$. For $T_{0}\ll 10^9\,\unit{K}$, the temperature is so low that the thermal electron-positron pair density decreases exponentially with temperature, leading to a very low optical depth.

\begin{figure}[H]
\centering
\includegraphics[angle=0,scale=.5]{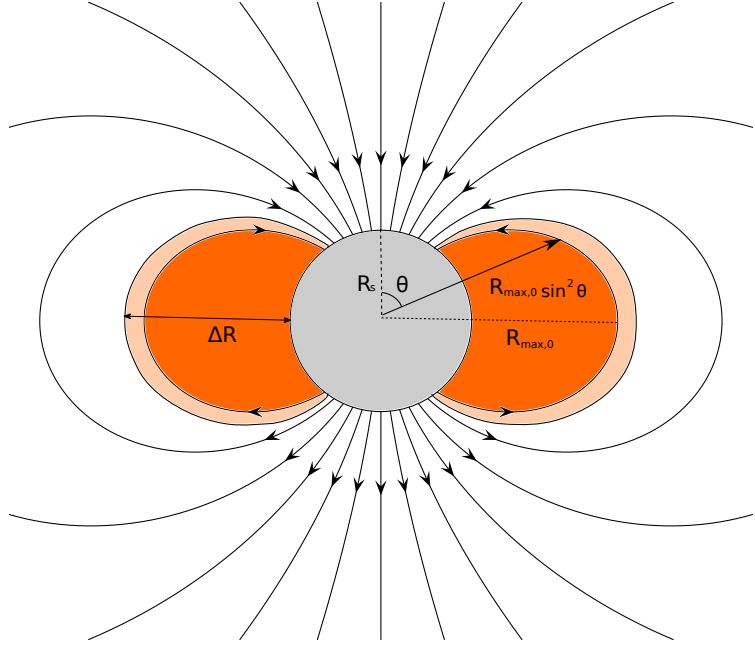}
\caption{The geometric structure of the trapped fireball. The grey circle represents the neutron star. The orange region represents the inner part of the trapped fireball. The thin bisque layer represents the atmosphere of the trapped fireball.}\label{fig1}
\end{figure}

\begin{figure}[H]
\centering
\includegraphics[angle=0,scale=.35]{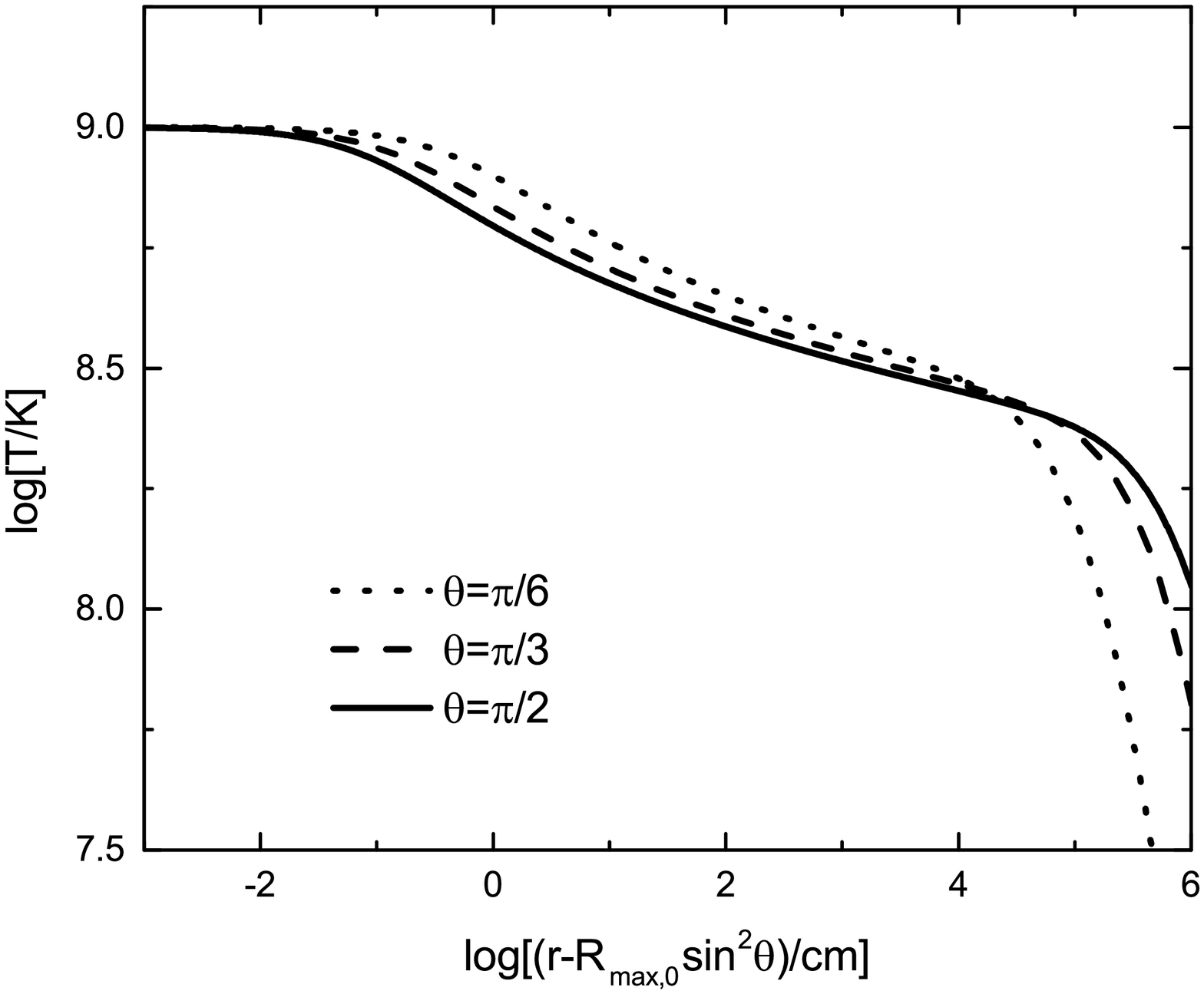}
\caption{The temperature as a function of the radial distance to the atmosphere bottom ($r-R_{max,0}\sin^2\theta$) for $B_p=10^{15}\,\unit{G}$, $R_{max,0}=2R_s$ and $T_0=10^9\,\unit{K}$. The dotted, dashed, and solid lines denote $\theta=\pi/6,\,\pi/3,\,\pi/2$, respectively. }\label{fig2}
\end{figure}

\begin{figure}[H]
\centering
\includegraphics[angle=0,scale=.35]{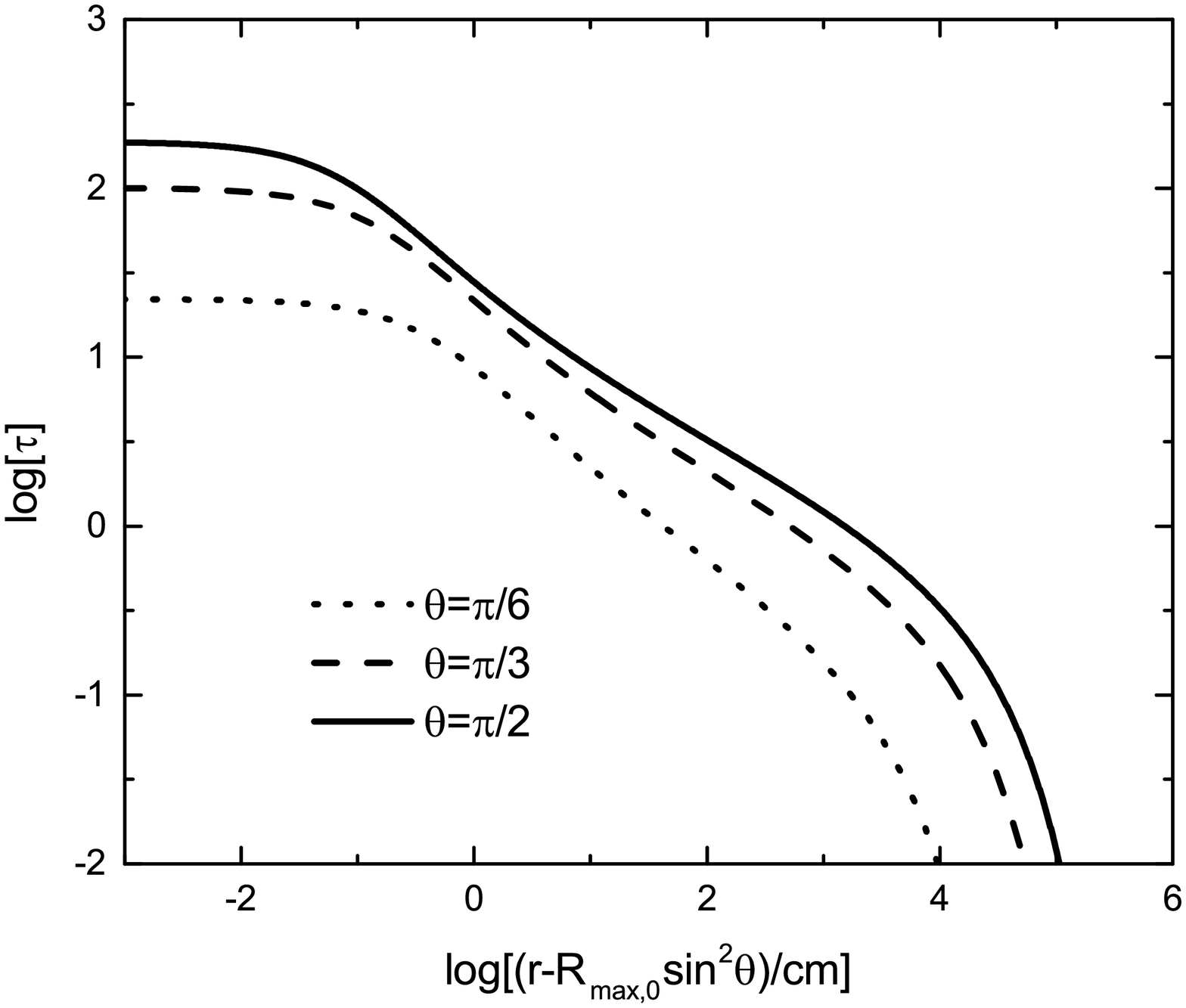}
\caption{The optical depth as a function of the radial distance to the atmosphere bottom ($r-R_{max,0}\sin^2\theta$) for $B_p=10^{15}\,\unit{G}$, $R_{max,0}=2R_s$ and $T_0=10^9\,\unit{K}$. The dotted, dashed, and solid lines denote $\theta=\pi/6,\,\pi/3,\,\pi/2$, respectively.}\label{fig3}
\end{figure}

\begin{figure}[H]
\centering
\includegraphics[angle=0,scale=.35]{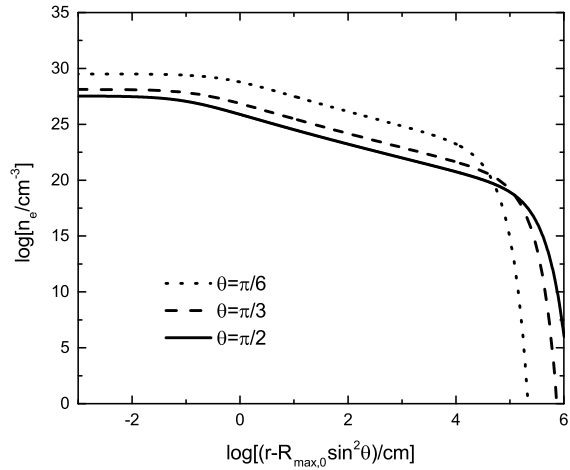}
\caption{The number density of electron-positron pairs as a function of the radial distance to the atmosphere bottom ($r-R_{max,0}\sin^2\theta$) for $B_p=10^{15}\,\unit{G}$, $R_{max,0}=2R_s$ and $T_0=10^9\,\unit{K}$. The dotted, dashed, and solid lines denote $\theta=\pi/6,\,\pi/3,\,\pi/2$, respectively.}\label{fig4}
\end{figure}

\begin{figure}[H]
\centering
\includegraphics[angle=0,scale=.35]{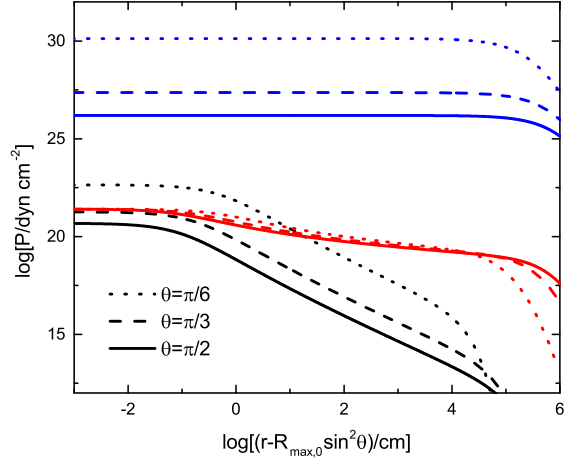}
\caption{The pressure as a function of the radial distance to the atmosphere bottom ($r-R_{max,0}\sin^2\theta$) for $B_p=10^{15}\,\unit{G}$, $R_{max,0}=2R_s$ and $T_0=10^9\,\unit{K}$. The dotted, dashed, and solid lines denote $\theta=\pi/6,\,\pi/3,\,\pi/2$, respectively. The electron-positron pairs, radiation and magnetic pressures are denoted by black, red and blue lines, respectively.}\label{fig5}
\end{figure}

In Fig. \ref{fig2}-\ref{fig5}, we plot the temperature, optical depth, number density of electron-positron pairs and pressure as a function of the radial distance from the bottom of the atmosphere  $r-R_{max,0}\sin^2\theta$, where $R_{max,0}\sin^2\theta$ is the radial distance from the neutron star center to the bottom of the atmosphere (see Fig. \ref{fig1}, with $\theta=\pi/6,\,\pi/3,\,\pi/2$ denoted by dotted, dashed and solid lines). For the given boundary condition described above, the atmosphere is very thin $\sim 10^3\,\unit{cm}$ and the effective temperature at the photosphere ($\tau=1$) is $T_{eff}=27\,\unit{keV}$ for $\theta=\pi/2$. This is consistent with the observed temperature of the pulsating tail. The temperatures for different $\theta$ have little difference for $\tau>1$, and the temperature near the magnetosphere equator is higher than that near the pole for $\tau<1$. The optical depth from the direction near the pole is lower than that in the direction of the equator, which means that the flux near the pole might be higher. As shown in Fig. \ref{fig4}, we find that the electron-positron pair number density decreases rapidly when $\tau<1$. Figure \ref{fig5} shows pressure as a function of the radial distance to the atmosphere bottom $r-R_{max,0}\sin^2\theta$. The electron-positron pairs, radiation and magnetic pressures are denoted by black, red and blue lines, respectively. One can see that the total pressure is dominated by the magnetic pressure, so that the pair plasma is trapped in the magnetosphere. Near the equator, one has $P_g<P_\gamma$. However, near the pole, the gas pressure could be larger than the radiation pressure at the atmosphere bottom. The result suggests that the pressure at the pole decreases more rapidly than that of the equator, since the temperature at the pole decreases from a smaller radius $r-R_{max,0}\sin^2\theta$.

\section{Photon Polarization and Scattering Opacity of Pairs}

\citet{ho03} calculated the photon polarization modes in a magnetized electron-ion plasma. In this section, we calculate the photon polarization modes in a magnetized electron-positron pair plasma following the procedure introduced by \citet{ho03}. As shown below, because of the charge symmetry in a pair plasma, some noticeable differences from the case of an electron-ion plasma exist. For example, the photons are completely linearly polarized, and there is no vacuum resonance point and ion cyclotron absorption in the case of the pair plasma.

For the atmosphere of a trapped fireball, the pair plasma is non-relativistic ($k_BT\ll m_ec^2$). In the coordinate system $x^\prime\,y^\prime\,z^\prime$ with $\mathbf{B}$ along $\mathbf{z}^\prime$, the dielectric tensor $\pmb{\epsilon}^{(p)}$ contributed by the plasma is given by \citep{ho03}
\begin{eqnarray}
\left[\pmb{\epsilon}^{(p)}\right]_{\mathbf{z}^\prime=\mathbf{\hat{B}}}=\left(
\begin{array}{ccc}
\varepsilon & ig & 0\\
-ig & \varepsilon & 0\\
0 & 0 & \eta\\
\end{array}
\right).
\end{eqnarray}
Here
\begin{eqnarray}
\varepsilon&=&1-\sum_s\frac{\lambda_s\upsilon_s}{\lambda_s^2-u_s},\\
\eta&=&1-\sum_s\frac{\upsilon_s}{\lambda_s},\\
g&=&-\sum_s\frac{\mathrm{sign}(q_s)u_s^{1/2}\upsilon_s}{\lambda_s^2-u_s},
\end{eqnarray}
where $s$ represents the charged particle species in the plasma, $u_s=\omega_{Bs}^2/\omega^2$, and $\upsilon_{s}=\omega_{ps}^2/\omega^2$. For the charged particle $s$, $\omega_{Bs}=|q_s|B/(m_sc)$ is the cyclotron frequency, and $\omega_{ps}=(4\pi n_sq_s^2/m_s)^{1/2}$ is the plasma frequency. The parameter $\lambda_s=1+i\nu_s/\omega$ delineates damping of the particle motion, where $\nu_s$ is the damping rate. In our calculation, we assume small damping ($\nu_s\ll \omega$). For electron-positron pairs, $s=e^+,e^-$, we have
\begin{eqnarray}
\varepsilon&=&1-\frac{2\upsilon_e}{1-u_e},\\
\eta&=&1-2\upsilon_e,\\
g&=&0,
\end{eqnarray}
where 
\begin{eqnarray}
u_e&=&\left(\frac{E_{Be}}{E}\right)^2=\left(\frac{1158B_{14}\,\mathrm{keV}}{E}\right)^2,\\
\upsilon_e&=&\left(\frac{E_{pe}}{E}\right)^2=\left(\frac{0.02871\rho_1^{1/2}\,\mathrm{keV}}{E}\right)^2,
\end{eqnarray}
with $B_{14}=B/(10^{14}\,\unit{G})$ and $\rho_1=\rho/(1\,\unit{g\,cm^{-3}})$. We note that the non-diagonal terms in the matrix are zero due to charge symmetry of pairs. On the other hand, in strong magnetic fields, vacuum polarization makes a corrected contribution to the dielectric tensor:
\begin{eqnarray}
\Delta\pmb{\epsilon}^{(v)}=(a-1)\mathbf{I}+q\mathbf{\hat{B}\hat{B}},
\end{eqnarray}
where $\mathbf{I}$ is the unit tensor and $\mathbf{\hat{B}}$ is the unit vector along $\mathbf{B}$. Similarly, vacuum polarization also modifies the magnetic permeability tensor $\pmb{\mu}$ as
\begin{eqnarray}
\mathbf{H}_w=\pmb{\mu}^{-1}\cdot\mathbf{B}_w=(a\mathbf{I}+m\mathbf{\hat{B}\hat{B}})\cdot\mathbf{B}_w,
\end{eqnarray}
where $\mathbf{H}_w$ and $\mathbf{B}_w$ are the magnetizing field and the magnetic field of an electromagnetic wave, respectively.

The vacuum polarization coefficients are given by \citep{har06}
\begin{eqnarray}
a&=&1-\frac{2\alpha}{9\pi}\ln\left(1+\frac{\beta_B^2}{5}\frac{1+0.25487\beta_B^{3/4}}{1+0.75\beta_B^{5/4}}\right),\\
q&=&\frac{7\alpha}{45\pi}\beta_B^2\frac{1+1.2\beta_B}{1+1.33\beta_B+0.56\beta_B^2},\\
m&=&-\frac{\alpha}{3\pi}\frac{\beta_B^2}{3.75+2.7\beta_B^{5/4}+\beta_B^2},
\end{eqnarray}
where $\beta_B\equiv B/B_Q$ is the relevant dimensionless magnetic field parameter, and $\alpha=1/137$ is the fine-structure constant.

If $|\epsilon^{(v)}_{ij}|\ll 1$ ($B\ll5\times10^{16}\mathrm{G}$), the vacuum polarization contribution, $\Delta\pmb{\epsilon}^{(v)}$, could be added linearly to the dielectric tensor $\pmb{\epsilon}^{(p)}$. In the frame with $\mathbf{\hat{B}}$ along $\mathbf{z}^\prime$, one has
\begin{eqnarray}
\left[\pmb{\epsilon}\right]_{\mathbf{z}^\prime=\mathbf{\hat{B}}}&=&\left(
\begin{array}{ccc}
\varepsilon & 0 & 0\\
0 & \varepsilon & 0\\
0 & 0 & \eta\\
\end{array}\right)+\left(
\begin{array}{ccc}
a-1 & 0 & 0\\
0 & a-1 & 0\\
0 & 0 & a+q-1\\
\end{array}
\right)\nonumber\\
&=&\left(
\begin{array}{ccc}
\varepsilon^\prime & 0 & 0\\
0 & \varepsilon^\prime & 0\\
0 & 0 & \eta^\prime\\
\end{array}\right),
\end{eqnarray}
where $\varepsilon^\prime=\varepsilon+a-1$ and $\eta^\prime=\eta+a+q-1$.

Using the Maxwell equations, the equation for a plane wave with $\mathbf{E}\propto \mathrm{e}^{i(\mathbf{k}\cdot\mathbf{r}-\omega t)}$ can be expressed as
\begin{eqnarray}
\nabla\times(\pmb{\mu}^{-1}\cdot\nabla\times\mathbf{E})=\frac{\omega^2}{c^2}\pmb{\epsilon}\cdot\mathbf{E},
\end{eqnarray}
which is \citep{ho03}, 
\begin{eqnarray}
\left\lbrace\frac{1}{a}\epsilon_{ij}+n^2\left[\hat{k_i}\hat{k_j}-\delta_{ij}-\frac{m}{a}(\hat{k}\times\hat{B})_i(\hat{k}\times\hat{B})_j\right]\right\rbrace E_j=0, \label{wave}
\end{eqnarray}
where $n=ck/\omega$ is the refractive index. In the coordinate system $x\,y\,z$ with $\mathbf{k}$ along $z$-axis and $\mathbf{B}$ in the $x-z$ plane, the dielectric tensor is given by
\begin{eqnarray}
\left[\pmb{\epsilon}\right]_{\mathbf{z}=\mathbf{\hat{k}}}=\left(
\begin{array}{ccc}
\varepsilon^\prime\cos^2\theta_B+\eta^\prime\sin^2\theta_B & 0 & (\varepsilon^\prime-\eta^\prime)\sin\theta_B\cos\theta_B\\
0 & \varepsilon^\prime & 0\\
(\varepsilon^\prime-\eta^\prime)\sin\theta_B\cos\theta_B & 0 & \varepsilon^\prime\sin^2\theta_B+\eta^\prime\cos^2\theta_B\\
\end{array}
\right),
\end{eqnarray}
where $\theta_B$ is the angle between $\mathbf{k}$ and $\mathbf{B}$. According to the $z$-component of Eq.(\ref{wave}), one has
\begin{eqnarray}
E_z=-(\epsilon_{zx}/\epsilon_{zz})E_{x}.
\end{eqnarray}
Then Eq.(\ref{wave}) becomes
\begin{eqnarray}
\left(
\begin{array}{cc}
\eta_{xx}-n^2 & 0\\
0 & \eta_{yy}-rn^2\\
\end{array}
\right)\left(
\begin{array}{c}
E_x\\
E_y\\
\end{array}
\right)=0,\label{marix}
\end{eqnarray} 
where $r\equiv1+(m/a)\sin^2\theta_B$, and
\begin{eqnarray}
\eta_{xx}&=&\frac{1}{a\epsilon_{zz}}(\epsilon_{zz}\epsilon_{xx}-\epsilon_{xz}\epsilon_{zx})=\frac{1}{a\epsilon_{zz}}\varepsilon^\prime\eta^\prime,\\
\eta_{yy}&=&\frac{1}{a\epsilon_{zz}}\epsilon_{zz}\epsilon_{yy}=\frac{1}{a\epsilon_{zz}}[(\varepsilon^{\prime2}-\varepsilon^\prime\eta^\prime)\sin^2\theta_B+\varepsilon^\prime\eta^\prime].
\end{eqnarray}
There are two solutions to Eq.(\ref{marix}):

(i.) If $n^2=\eta_{xx}$, then $E_y=0$ and $E_x,E_z\neq0$. The electric field vector of the wave is in the $\mathbf{k-B}$ plane. This is the ordinary mode (O mode). 

(ii.) If $n^2=\eta_{yy}/r$, then $E_x=E_z=0$ and $E_y\neq0$. The electric field vector of the wave is perpendicular to the $\mathbf{k-B}$ plane. This is the extraordinary mode (E mode).

Different from the electron-ion plasma, we find that for electron-positron pairs, one does not need to define the polarization ellipticity $K_j=-iE_x/E_y$ \citep{ho03}, and both modes are linearly polarized. In this case, there is no ``vacuum resonance" point where E mode vs. O mode classification becomes ambiguous and can convert from one mode to the other.

The mode eigenvector in the coordinate system with the wave vector  $\mathbf{k}$ along $z$-axis is
\begin{eqnarray}
\pmb{e}^\prime=\frac{1}{(E_x^2+E_y^2+E_z^2)^{1/2}}(E_x,E_y,E_z).
\end{eqnarray} 
The scattering opacity depends on the polarization vector through its projection on the coordinate frame with the $z$-axis along the magnetic field $\mathbf{B}$ direction \citep{ho01,ho03}. In this new coordinate system with $\mathbf{B}$ along $z$-axis, the eigenvector is given by
\begin{eqnarray}
\pmb{e}=\frac{1}{(E_x^2+E_y^2+E_z^2)^{1/2}}(E_x\cos\theta_B+E_z\sin\theta_B,E_y,-E_x\sin\theta_B+E_z\cos\theta_B).
\end{eqnarray}
The cyclic components of $\pmb{e}$ read\footnote{In \citet{ho01,ho03}, they defined $iK_j=E_x/E_y$ and $iK_{z,j}=E_z/E_y$, and $K_j$ and $K_{z,j}$ are complex numbers, which are determined by the plasma medium properties. Thus, there is a plus or minus sign in $|{e_\pm^j}|^2$ in their notation. }
\begin{eqnarray}
|{e_\pm^j}|^2&=&\left|\frac{1}{\sqrt 2}({e_x^j}\pm i {e_y^j})\right|^2=\frac{|E_y|^2+|E_x\cos\theta_B+E_z\sin\theta_B|^2}{2(E_x^2+E_y^2+E_z^2)},\nonumber\\
|{e_0^j}|^2&=&|{e_z^j}|^2=\frac{|E_x\sin\theta_B-E_z\cos\theta_B|^2}{E_x^2+E_y^2+E_z^2},
\end{eqnarray}
where $j=1$ corresponds to the E mode, and $j=2$ corresponds to the O mode. Specifically, for the E mode, the cyclic components are given by
\begin{eqnarray}
|e_\pm^1|^2=\frac{1}{2},\,|e_0^1|^2=0, \label{base}
\end{eqnarray}
and for the O mode, they are given by
\begin{eqnarray}
|e_\pm^2|^2=\frac{|\cos\theta_B+K_{zx}\sin\theta_B|^2}{2(1+K_{zx}^2)},\nonumber\\
|e_0^2|^2=\frac{|\sin\theta_B-K_{zx}\cos\theta_B|^2}{1+K_{zx}^2},
\end{eqnarray}
where we define $K_{zx}\equiv E_z/E_x=-(\epsilon_{zx}/\epsilon_{zz})$. We note that the eigenvectors of the E mode are constants for the electron-positron plasma. This is different from the case of electron-ion plasma. The scattering opacity from the $j$ mode into the $i$ mode is given by \citep{ven79,ho01,ho03}
\begin{eqnarray}
\kappa_{ji}^{sc}=\frac{\sigma_T}{m_e}\sum_{\alpha=-1}^1\left[(1+\alpha u_e^{1/2})^2+\gamma_e^2\right]^{-1}|e_\alpha^j|^2A_\alpha^i,
\end{eqnarray}
where $\gamma_e\equiv\nu_e/\omega$ is the damping factor, and $A_\alpha^i=(3/4)\int_0^\pi d\theta \sin\theta |e_\alpha^i|^2$. The electron scattering opacity from the mode $j$ (into both model $i$ and mode $j$) is
\begin{eqnarray}
\kappa_{j}^{sc}=\frac{\sigma_T}{m_e}\sum_{\alpha=-1}^1\left[(1+\alpha u_e^{1/2})^2+\gamma_e^2\right]^{-1}|e_\alpha^j|^2A_\alpha,\label{opacji}
\end{eqnarray}
where $A_\alpha=A_\alpha^1+A_\alpha^2$. The scattering opacities for the two modes can be approximately derived \citep{mes92,ulm94,mil95}. For the E-mode, one has
\begin{eqnarray}
\kappa_1^{sc}=\kappa_{11}^{sc}+\kappa_{12}^{sc},\,\,\kappa_{11}^{sc}\simeq 3\kappa_{12}^{sc},\label{scaE}
\end{eqnarray}
where the second equation is under the condition of $\omega m_ec/eB\sim10^{-2}-10^{-3}$; For the O-mode, one has
\begin{eqnarray}
\kappa_2^{sc}=\kappa_{22}^{sc}+\kappa_{21}^{sc},\,\,\kappa_{21}^{sc}\simeq \left(\frac{\omega}{\omega_{Be}}\right)^2\kappa_{22}^{sc}\ll\kappa_2^{sc}. \label{scaO}
\end{eqnarray}
According to Eq.(\ref{scaE})-Eq.(\ref{scaO}), the probability of E-mode photons converting into O-mode photons is $P_{EO}\simeq 1/4$, and the probability of O-mode photons converting into E-mode photons is $P_{OE}\simeq (\omega/\omega_{Be})^2$.
For the transverse-mode approximation ($K_{zx}\ll1$), one has $A_\alpha\simeq1$. Thus, the opacity is approximately
\begin{eqnarray}
\kappa_{1}^{sc}\simeq\frac{\sigma_T}{2m_e}\left[\frac{\omega^2}{(\omega-\omega_{Be})^2+(\gamma_e\omega)^2}+\frac{\omega^2}{(\omega+\omega_{Be})^2+(\gamma_e\omega)^2}\right]
\end{eqnarray}
for the E mode, and 
\begin{eqnarray}
\kappa_{2}^{sc}\simeq\frac{\sigma_T}{2m_e}\left[\frac{\omega^2\cos^2\theta_B}{(\omega-\omega_{Be})^2+(\gamma_e\omega)^2}+\frac{\omega^2\cos^2\theta_B}{(\omega+\omega_{Be})^2+(\gamma_e\omega)^2}+\frac{2\sin^2\theta_B}{1+\gamma_e^2}\right]
\end{eqnarray}
for the O model. For radiation frequencies below the cyclotron frequency, i.e. $\omega\ll\omega_{Be}$, and the damping factor $\gamma_e\ll 1$, one has
\begin{eqnarray}
\kappa_1^{sc}\simeq\frac{\omega^2}{\omega_{Be}^2}\kappa_T,\,\,\kappa_2^{sc}\simeq\left(\frac{\omega^2}{\omega_{Be}^2}\cos^2\theta_B+\sin^2\theta_B\right)\kappa_T, \label{opacity}
\end{eqnarray}
where $\kappa_T = \sigma_T/ m_e$ ($\sigma_T$ is the Thomson cross section, and $m_e$ is the electron mass). Due to the absence of the ion cyclotron absorption and vacuum resonance absorption in the case of electron-positron plasma, the above equations could be approximately applied to calculate the opacities for $\omega\ll\omega_{Be}$. We then finally obtain the Rosseland mean opacity $\kappa$ through
\begin{eqnarray}
\frac{1}{{\kappa}}=\left[\int_0^\infty\left(\frac{1}{\kappa_1^{sc}}+\frac{1}{\kappa_2^{sc}}\right)\frac{\partial B_\nu(T)}{\partial T}d\nu\right]\Bigg/\left[\int_0^\infty\frac{\partial B_\nu(T)}{\partial T}d\nu\right].
\end{eqnarray}
 
For electron energy $\lesssim300\mathrm{keV}$, the contribution from electron-electron bremsstrahlung can be safely ignored \citep{hau75}. We therefore ignore free-free absorption of electron pairs in the atmosphere of the trapped fireball.

\begin{figure}[H]
\centering
\includegraphics[angle=0,scale=.3]{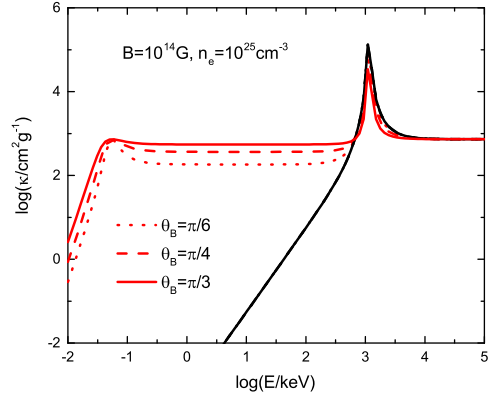}
\caption{The scattering opacities $\kappa^{sc}$ as a function of energy for various angles between $\mathbf{k}$ and $\mathbf{B}$, $\theta_B$, for $B=10^{14}\,\unit{G}$ and $n_e=10^{25}\,\unit{cm^{-3}}$. The black curves represent the E model opacity, and the red curves represent the O mode opacity. The dotted, dashed and solid curves denote $\theta_B=\pi/6,\,\pi/4,\,\pi/3$, respectively.} \label{fig6}
\end{figure}

\begin{figure}[H]
\centering
\includegraphics[angle=0,scale=.3]{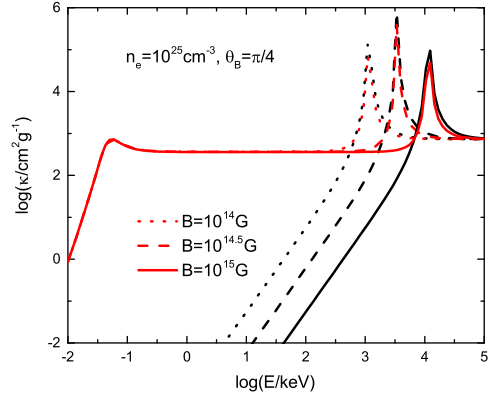}
\caption{The scattering opacities $\kappa^{sc}$ as a function of energy for different magnetic field strengths $B$ for $n_e=10^{25}\,\unit{cm^{-3}}$ and $\theta_B=\pi/4$. The black curves represent the E mode opacity, and the red curves represent the O mode opacity. The magnetic field strengths $B=10^{14},\,10^{14.5},\,10^{15},\,\unit{G}$ are denoted by dotted, dashed and solid curves, respectively.} \label{fig7}
\end{figure}

\begin{figure}[H]
\centering
\includegraphics[angle=0,scale=.3]{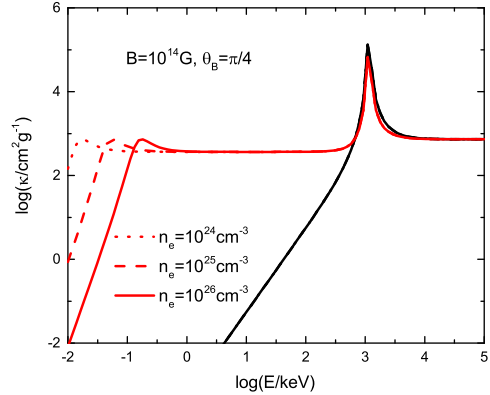}
\caption{The scattering opacities $\kappa^{sc}$ as a function of energy for various electron-positron pair number density $n_e$ for $B=10^{14}\,\unit{G}$ and $\theta_B=\pi/4$. The black curves represent the E mode opacity, and the red curves represent the O mode opacity. The pair number densities $n_e=10^{24},\,10^{25},\,10^{26}\,\unit{cm^{-3}}$ are denoted by dotted, dashed and solid curves, respectively.} \label{fig8}
\end{figure}

We apply the full expressions of $\kappa_1^{sc}$ and $\kappa_2^{sc}$ in Eq. (\ref{opacji}) to perform the calculation. The scattering opacities $\kappa_j^{sc}$ of the two modes as a function of energy for various angle $\theta_B$, magnetic fields $B$, and electron-positron pair density $n_e$ are presented in Figures \ref{fig6} - \ref{fig8}, respectively. Different from the case of electron-ion plasma \citep{ho01,ho03}, for electron-positron pair plasma, there are no ion cyclotron absorption and vacuum resonance absorption, so that the opacities satisfy Eq. (\ref{opacity}) for $E\ll E_{Be}=\hbar eB/m_ec$. As shown in Fig.\ref{fig6}, for magnetic field $B=10^{14}\,\unit{G}$ and electron-positron pair number density $n_e=10^{25}\,\unit{cm^{-3}}$, the O mode opacity exhibits an angle dependence for $\theta_B$, i.e. $\kappa_O\simeq\kappa_T\sin^2\theta_B\,(\omega\ll\omega_{Be})$. The E model opacity, on the other hand, is independent of $\theta_B$ (see Eq. (\ref{base}) and Eq. (\ref{opacity})). As shown in Fig \ref{fig7}, for $n_e=10^{25}\,\unit{cm^{-3}}$ and $\theta_B=\pi/4$, the electron cyclotron peak $E_{Be}$ is proportional to the strength of magnetic field $B$ and the E mode opacity satisfies $\kappa_E\simeq(\omega/\omega_{Be})^2\kappa_T$ for $\omega\ll\omega_{Be}$. As shown in Fig. \ref{fig8}, for $B=10^{14}\,\unit{G}$ and $\theta_B=\pi/4$, the plasma peak  $E_{pe}$ is proportional to $n_e^{1/2}$.

\section{Monte Carlo Simulations}

\begin{figure}
\centering
\includegraphics[angle=0,scale=.5]{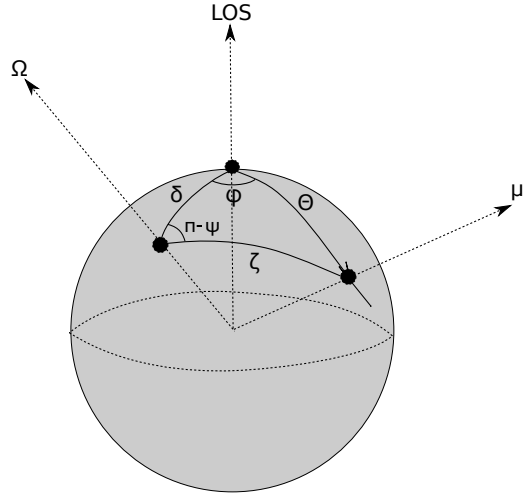}
\caption{The angles and vectors introduced in Section 4.}\label{fig9}
\end{figure}

In this section, we conduct Monte Carlo simulations to calculate the polarization properties of the X-ray photons emitted from the trapped fireball.
To calculate the polarization signals, we set a coordinate system\footnote{Notice that the coordinate systems $XYZ$ and $X'Y'Z'$ introduced in this section are the global coordinate systems for the neutron star, which are very different from the ``local" coordinate systems $xyz$ and $x'y'z'$ introduced in Section 3. In the following derivations, the coordinate symbols are still kept in lower case for these global coordinate systems.} $XYZ$ with $Z$-axis along the line of sight $\pmb{e_z}$, as shown in Fig. \ref{fig9}. The spin angular velocity vector $\pmb{\Omega}$ is in the $XZ$ plane. We define $\zeta$ as the angle between the magnetic dipole axis $\pmb{\mu}$ and $\pmb{\Omega}$, $\Theta$ as the angle between $\pmb{\mu}$ and the line of sight $\pmb{e_z}$, and $\delta$ as the angle between $\pmb{e_z}$ and $\pmb{\Omega}$.

For a dipolar field, in the coordinate system $X^\prime Y^\prime Z^\prime$ with $\pmb{\mu}$ along the $Z^\prime$-axis, the magnetic field components in a spherical coordinate system are given by
\begin{eqnarray}
B_r^\prime&=&\frac{B_pR_s^3}{r^3}\cos\theta,\\
B_\theta^\prime&=&\frac{B_pR_s^3}{2r^3}\sin\theta,\\
B^\prime(r,\theta)&=&\frac{B_p}{2}\left(\frac{R_s}{r}\right)^3\sqrt{1+3\cos^2\theta}.
\end{eqnarray}
Transforming from the spherical coordinate system to the Cartesian coordinate system, we have
\begin{eqnarray}
B_x^\prime&=&\frac{B_pR_s^3}{2r^3}3\sin\theta\cos\theta\cos\varphi,\\
B_y^\prime&=&\frac{B_pR_s^3}{2r^3}3\sin\theta\cos\theta\sin\varphi,\\
B_z^\prime&=&\frac{B_pR_s^3}{2r^3}(2\cos^2\theta-\sin^2\theta),
\end{eqnarray}
where
\begin{eqnarray}
r&=&\sqrt{x^{\prime2}+y^{\prime2}+z^{\prime}},\\
\tan\theta&=&\sqrt{x^{\prime2}+y^{\prime2}}/z^{\prime},\\
\tan\varphi&=&y^\prime/x^\prime.
\end{eqnarray}
In the coordinate system $XYZ$ with the line of sight along the $Z$-axis, the magnetic field components can be calculated by
\begin{eqnarray}
\left(
\begin{array}{c}
B_x\\
B_y\\
B_z\\
\end{array}
\right)=
\left(
\begin{array}{ccc}
\cos\Theta\cos\phi & \sin\phi & -\sin\Theta\cos\phi\\
-\cos\Theta\sin\phi & \cos\phi & \sin\Theta\sin\phi\\
\sin\Theta & 0 & \cos\Theta\\
\end{array}
\right)
\left(
\begin{array}{c}
B_x^\prime\\
B_y^\prime\\
B_z^\prime\\
\end{array}
\right),
\end{eqnarray}
and the coordinate system transformation is given by
\begin{eqnarray}
\left(
\begin{array}{c}
x^\prime\\
y^\prime\\
z^\prime\\
\end{array}
\right)=
\left(
\begin{array}{ccc}
\cos\Theta\cos\phi & -\cos\Theta\sin\phi & \sin\Theta\\
\sin\phi & \cos\phi & 0\\
-\sin\Theta\cos\phi & \sin\Theta\sin\phi & \cos\Theta\\
\end{array}
\right)
\left(
\begin{array}{c}
x\\
y\\
z\\
\end{array}
\right),
\end{eqnarray}
where 
\begin{eqnarray}
\tan\phi&=&\frac{\sin\zeta\sin\psi}{\cos\zeta\sin\delta+\sin\zeta\cos\delta\cos\psi},\\
\cos\Theta&=&\cos\zeta\cos\delta-\sin\zeta\sin\delta\cos\psi,\\
\sin\Theta&=&\frac{\sin\zeta\sin\psi}{\sin\phi},
\end{eqnarray}
with $\phi$ being the azimuthal angle, and $\psi=\Omega t+\psi_0$ being the rotation phase ($\psi_0$ is the initial rotation phase). We assume that the photon flux is same everywhere at the bottom of the atmosphere. In order to calculate the photon spatial distribution, we need to know the surface area of the magnetosphere $S(\theta)$ with $\theta$. Because the border of the magnetosphere is given by $r=R_{max}\sin^2\theta$, where $R_{max}$ is the maximum radius of the closed field lines, the length of the dipole field line is given by
\begin{eqnarray}
dl&=&\sqrt{(rd\theta^\prime)^2+(dr)^2}\nonumber\\
&=&-R_{max}\sqrt{1+3\cos^2\theta}d(\cos\theta).
\end{eqnarray}
Thus, the differential area of the magnetosphere is given by
\begin{eqnarray}
dS&=&2\pi r\sin\theta dl\nonumber\\
&=&2\pi R_{max}^2\sin^4\theta\sqrt{1+3\cos^2\theta}d\theta.\label{ds}
\end{eqnarray}
Integrating over $dS$, we obtain the area of the closed magnetic field line within $\theta$, i.e. $S(\theta)=\int_0^\theta dS$. The total area is $S(\theta=\pi)\simeq8.9R_{max}^2$.
Using this information, we can apply the Monte Carlo method to assign the initial spatial distribution of the photons.

As we have discussed above, the observations show that the spectra of the pulsating tails are dominated by the thermal component. Therefore, in this paper, we only consider the thermal component contributed by the photon-pair trapped fireball. Because the optical depth of the trapped fireball is very large, the observed radiation comes from the atmosphere of the trapped fireball. We apply the atmosphere model described in Sec. 2. to calculate the photon polarization properties through Monte Carlo simulations. The procedure of our calculations is the following:

A. We assume that the boundary of the atmosphere of the trapped fireball satisfies the follow conditions:

(i.) The bottom of the atmosphere is a closed magnetic field line surface defined by $r=R_{max,0}\sin^2\theta$ with a same temperature $T_0$.

(ii.) The radiative flux is the same everywhere at the bottom of the atmosphere. As a result, the initial photon spatial distribution can be simulated according to the coordinates $(\theta,\phi)$, i.e. $\xi=\int_0^\theta dS/\int_0^\pi dS$, $dS$ is given by Eq. (\ref{ds}), $\phi=2\pi\xi$, with $\xi$ being a random number between (0,1). The initial directional distribution $(\theta_v,\phi_v)$ of the photons is isotropic, i.e., $\cos\theta_v=1-2\xi$ and $\phi_v=2\pi\xi$. 

(iii.) The initial photon number in the E mode is equal to that in the O mode at the bottom of the atmosphere. However, the photons in one mode can be converted to the other mode due to the electron scattering in the atmosphere, leading to the change of the ratio between two modes.

B. We keep track the propagation and transition of each photon through the medium.

(i.) At each time step, we grid the atmosphere of the trapped fireball in 3 dimensions. The properties (e.g. density, temperature, pressure) in each cell are kept constant.

(ii.) The scattering optical depth $\tau_{sc}$ is given by $\tau_{sc}=-\ln(1-\xi)$. Each photon is allowed to travel along its trajectory in each time step. The optical depth in each cell is calculated until $\tau>\tau_{sc}$. Note that we need to use different opacities for different photon modes in this process.

(iii.) Scattering would change the mode of the photons. The probability for E mode converting into O mode is $P_{EO}\simeq1/4$, whereas the probability for O mode converting into E-mode is $P_{OE}\simeq(\omega/\omega_{Be})^2$.

(iv.) Due to the electrons' thermal motion, scattering would be isotropic. The direction $(\theta,\phi)$ of a scattered photon is given by $\cos\theta_v=1-2\xi$ and $\phi_v=2\pi\xi$.

(v.) We assume that the atmosphere satisfies local thermodynamic equilibrium (LTE). The energy distribution of the photons depends on the energy distribution of the electron-positron pairs that have a thermal distribution.

(vi.) We repeat the above procedure for each time step. Based on the atmosphere model discussed above, we find that the photons are almost no longer scattered when $r\gtrsim10\,R_s$, since $\tau(r=10R_s)\ll 1$.

C. We collect the photons along the light of sight within the solid angel $\Delta\Omega=0.15$, and calculate the directional angel $\phi_B$ for both photon modes at $r=10R_s$. For E-mode photons, one has $\phi_{B,E}=\phi_B$, where
\begin{eqnarray}
\phi_B\simeq\left \{
\begin{array}{lll}
\arctan(B_y/B_x), &\,\,& B_x>0,\\
\pi/2+\arctan(B_y/B_x), &\,\,& B_x<0,\\
\end{array}
\right.
\end{eqnarray}
and $B_x$ and $B_y$ are the magnetic field components perpendicular to the light of sight. For O mode photons, one has $\phi_{B,O}=\pi/2+\phi_B$.

The Stokes parameters near the photosphere are given by
\begin{eqnarray}
I&=&\langle E_x^2\rangle+\langle E_y^2\rangle,\\
Q&=&\langle E_x^2\rangle-\langle E_y^2\rangle,\\
U&=&\langle E_a^2\rangle-\langle E_b^2\rangle,\\
V&=&\langle E_l^2\rangle-\langle E_r^2\rangle,
\end{eqnarray}
where the subscripts refer to different bases of the Jones vector space: $(x,y)$ is the Cartesian basis, $(a,b)$ is the Cartesian basis rotated by $\pi/4$, $(l,r)$ is the circular basis, which is defined as $e_{l,r}=(e_x\pm ie_y)/\sqrt{2}$, and $\langle E_x^2\rangle$, $\langle E_y^2\rangle$, $\langle E_a^2\rangle$ and $\langle E_b^2\rangle$ are given by
\begin{eqnarray}
\langle E_x^2\rangle&=&\sum_i\cos^2\phi_{B,\alpha_i},\\
\langle E_y^2\rangle&=&\sum_i\sin^2\phi_{B,\alpha_i},\\
\langle E_a^2\rangle&=&\frac{1}{2}\sum_i(\cos\phi_{B,\alpha_i}-\sin\phi_{B,\alpha_i})^2,\\
\langle E_b^2\rangle&=&\frac{1}{2}\sum_i(\cos\phi_{B,\alpha_i}+\sin\phi_{B,\alpha_i})^2,
\end{eqnarray}
where $i$ presents each observed photon and $\alpha_i$ presents the mode of the photon $i$.
Finally, the polarization degree near the photosphere is given by
\begin{eqnarray}
\Pi=\frac{\sqrt{Q^2+U^2}}{I},
\end{eqnarray}
and the polarization angle is
\begin{eqnarray}
\chi=\frac{1}{2}\arctan\left(\frac{U}{Q}\right).
\end{eqnarray}

After a photon escapes the trapped fireball, it will propagate in the magnetic vacuum. The dielectric and inverse permeability are $\pmb{\epsilon}=a\mathbf{I}+q\mathbf{\hat{B}\hat{B}}$ and $\pmb{\mu}^{-1}=a\mathbf{I}+m\mathbf{\hat{B}\hat{B}}$, respectively. For a transverse electromagnetic wave emitted from a point in the magnetosphere atmosphere, it can be treated as the superposition of E-mode and O-mode photons, i.e.
\begin{eqnarray}
\mathbf{E}=A_O\pmb{e_O}+A_E\pmb{e_E},
\end{eqnarray}
where
\begin{eqnarray}
\pmb{e_O}=(\cos\phi,\sin\phi),\,\pmb{e_E}=(-\sin\phi,\cos\phi)
\end{eqnarray}
with the refraction indices for different modes being
\begin{eqnarray}
n_O\simeq1+(q/2)\sin^2\theta_B,\,n_E\simeq1-(m/2)\sin^2\theta_B.
\end{eqnarray}
One therefore has $\Delta n\equiv n_O-n_E=(1/2)(q+m)\sin^2\theta_B$. The evolution of the mode amplitudes is given by \citep{lai02}
\begin{eqnarray}\label{eq:lai-ho}
i\frac{\mathrm{d}}{\mathrm{d} s}\left(
\begin{array}{c}
A_O\\
A_E\\
\end{array}
\right)=
\left(
\begin{array}{cc}
-(\omega/c)\Delta n/2 & i\mathrm{d}\phi/\mathrm{d} s\\
-i\mathrm{d}\phi/\mathrm{d} s & (\omega/c)\Delta n/2\\
\end{array}
\right)
\left(
\begin{array}{c}
A_O\\
A_E\\
\end{array}
\right),
\end{eqnarray}
where $s=ct$ represents the photon trajectory. The condition for adiabatic evolution of the photon modes is 
\begin{equation}\label{eq:adiabatic}
(\omega/c)\Delta n\gg2\mathrm{d}\phi/\mathrm{d} s, 
\end{equation}
so that the non-diagonal elements in the matrix of Eq.(\ref{eq:lai-ho}) is essentially zero, and the normal modes do not mix \citep{hey00}. As the photon propagates in the magnetosphere, the QED correction to the dielectric tensor falls off as $B^2\sim r^{-6}$, and non-diagonal components would appear. Within the neutron star atmosphere context, \citet{hey02} first defined the QED polarization limiting radius $r_p$ (the distance from the star at which the adiabatic approximation breaks down), and found that the QED limit radius allows the neutron star surface polarization to be sufficiently large to be observed \citep[see also][]{hey03}. Here we perform a similar treatment for the trapped fireball. We approximately treat the adiabatic region to have a sharp edge at $r_p$. Based on the adiabatic condition (\ref{eq:adiabatic}), we calculate the polarization-limit radius, $r_p$, which is set by the condition $\omega\Delta n/c=2\mathrm{d}\phi/\mathrm{d} s$, i.e.\citep{ade06}
\begin{eqnarray} \label{eq:QED-radius}
r_p=153\,R_s f_\phi\left(\frac{E}{1\unit{keV}}\right)^{1/6}\left(\frac{B}{10^{14}\unit{G}}\right)^{1/3}\left(\frac{2\pi}{\Omega}\right)^{1/6},
\end{eqnarray}
where $R_s$ is the neutron star radius, $f_\phi$ is a slowly varying function of phase and is of the order of unity, $E$ is the photon energy, $B$ is the strength of magnetic filed, and $\Omega$ is the spin angular velocity. One can see that the adiabatic condition is safely satisfied in the trapped fireball and its atmosphere.
Beyond $r_p$, the magnetic field strength has dropped enough so that it no longer affects the state of photon polarization, i.e. the photon polarization state is `frozen'. Because $r_p\gg 10R_s$, the electric vectors of all the photons with a certain energy $E$ emitted from the photosphere would rotate for a same angle $\sim\phi(r_p)$, so that the polarization degree is no longer affected.

Since magnetars are slow rotators, the effect of magnetar spin is not important in the polarization calculations. For simplicity, in the Monte Carlo simulations, we assume an aligned rotator, i.e. 
the spin angular velocity vector $\pmb{\Omega}$ is parallel to the magnetic axis $\pmb{\mu}$, so that $\zeta=0$. All the photons with a certain energy have almost the same rotation angle $\phi(r_p)$ under the adiabatic condition. In this case, the Stokes parameter $U=0$, so that the polarization degree is $\Pi=Q/I$ and the polarization angle is $\chi=0$. We consider two energy bands, i.e. a soft band ($1-30\,\unit{keV}$) and a hard band ($30-100\,\unit{keV}$), and collect 5000 photons in each band within a solid angel $\Delta\Omega=0.15$ along the light of sight. The E mode and O mode are defined by the direction of the magnetic field at $r=10R_s$, where the adiabatic condition is satisfied.

\begin{figure}
\centering
\includegraphics[angle=0,scale=.5]{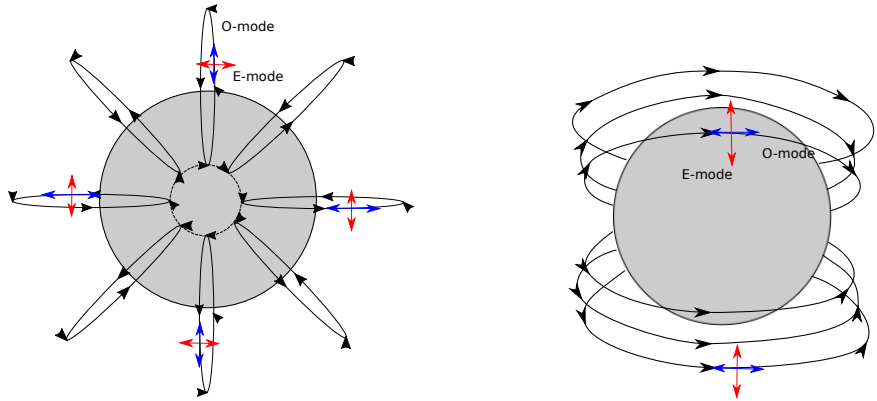}
\caption{Schematic pictures of the magnetar magnetosphere. The grey circle represents the neutron star. The black curves represent the magnetic field lines.
The red arrows represent the electric vectors of the E-mode photons, whereas the blue arrows represent the electric vectors of the O-mode photons. The left panel corresponds to the case with the line of sight parallel to the magnetic axis, whereas the right panel corresponds to the case with the line of sight perpendicular to the magnetic axis.}\label{fig10}
\end{figure}

Our simulation results are displayed in Table \ref{tab1}. As shown in the table, there are more E-mode photons in the soft band than in the hard band, since the opacity of the E-mode photons $\kappa_E\propto\omega^2$. The number ratio between the E-mode and O-mode photons, $N_E/N_O$, is higher for $\Theta=0$ than that for $\Theta=\pi/2$. This is because the opacity of E-mode photons is suppressed in strong magnetic fields, i.e. $\kappa_E\propto B^{-2}$, and $B$ is stronger near the polar. On the other hand, the polarization degree is close to zero for $\Theta=0$ and reaches the maximum value at $\Theta=\pi/2$, i.e. 27.9\% in the 1-30 keV band and 10.0\% in the 30-100 keV band. The result could be qualitatively explained from the direction of the magnetic field and the photon number ratio between E-mode and O-mode. As shown in the left panel of Fig. \ref{fig10}, if the light of sight is parallel to the magnetic axis, the directional angles of E-mode and O-mode are evenly distributed from 0 to $\pi$. Therefore, even though $N_E/N_O$ has the maximum value, the polarization degree is still zero. On the other hand, as shown in the right panel of Fig. \ref{fig10}, if the light of sight is perpendicular to the magnetic axis, the projected directions of the magnetic field lines almost have the same direction. In this case, the polarization degree depends on $N_E/N_O$. Since $N_E/N_O$ is of the order of unity, the polarization degree is mainly determined by the direction of the magnetic fields, so that an ordered $B$ field configuration naturally results in a relatively large polarization degree.

\section{Conclusions and Discussions}

Polarization observations provide the key information to probe the structure of the magnetar magnetosphere during giant flares. After the initial spike, the residual energy would form an optically thick photon-pair plasma. This photon-pair plasma is trapped in the magnetar magnetosphere, forming a ``trapped fireball''. Observations show that the flux of the thermal component is much higher than that of the non-thermal component between $10\,\unit{keV}$ to $100\,\unit{keV}$ in the early and later stages \citep{bog07}. In this paper, we assumed that the trapped fireball is enclosed by a set of closed-field lines in the magnetosphere, and calculated the atmospheric structure and polarization degree of the trapped fireball. We have reached the following conclusions:

A. Regarding the atmosphere of the trapped fireball:

(i.) The atmosphere of the trapped fireball is much thinner than the thickness of the trapped fireball. The size of a trapped fireball is a few times of $R_s$, which is determined by the requirement of the confinement of fireball energy within the magnetosphere, as shown in Eq. (\ref{trap}). Observations show that the effective radius of the thermal component is a few times of the radius of neutron star, which is consistent with this requirement. We calculated the structure of the trapped fireball and its atmosphere. Our results show that the thickness of the atmosphere is $\sim0.001R_s$ with a sharp temperature gradient.  

(ii.) In the atmosphere, the electron-positron pairs are non-relativistic, and the thermal pair density decreases with temperature (see Eq. (\ref{eos})). The pair gas pressure is lower than the radiative pressure at the photosphere, and the total pressure is dominated by the magnetic pressure. The atmosphere could be still in hydrostatic equilibrium due to the magnetic confinement. In the trapped fireball, gravity is not the dominant confining force, and the plasma is trapped due to the suppression of the transport of charged particles across the magnetic field lines.

(iii.) Opacity plays an important role in the structure of the trapped fireball. Different from the surface of a neutron star \citep{ho01,ho03}, the trapped fireball is dominated by the electron-positron pairs \citep{tho95,tho01}. We assume that the atmosphere is baryon free. For an electron-ion plasma, there is a vacuum resonance point where E-mode and O-mode photons could convert from one mode to another, and there are two peaks in the opacity-energy relation, e.g. the vacuum resonance absorption and the proton cyclotron absorption. However, for electron-positron pairs, both E-mode and O-mode photons are entirely linearly polarized. In this case, there is no vacuum resonance point where they can change mode during their propagation in the medium. Rather, for electron-positron pairs, there is only one electron cyclotron absorption peak with energy $E\simeq1158B_{14}\unit{keV}$, which is much larger than that of the observed effective temperature $k_BT_{\rm eff}\sim 30\,\unit{keV}$.

(iv.) The atmosphere temperature distribution is determined by the structure of the magnetosphere. For a specific angle $\theta$, the longer the distance from the magnetar, the lower the temperature. For a specific distance $r$, the temperature of the equator is higher than that of the pole (Note that the abscissa in Fig. \ref{fig2}-\ref{fig5} is the logarithm of $r-R_{max,0}\sin^2\theta$ rather than $r$). The pair density and opacity are determined by the temperature, so that the photosphere is no longer a sphere. Observationally one could only infer the effective size of the trapped fireball, but not the shape of the trapped fireball.  

(v.) Our results show that the opacity of E mode is independent of $\theta_B$ for electron-positron pairs, which is different from the case of the electron-ion plasma \citep{ho01,ho03}.

B. Regarding the polarization properties of the trapped fireball:

(i.) The opacity of the E-mode photons scales as $\kappa_E\propto\omega^2$. As a result, the number of E-mode photons is larger than that of O-mode photons in softer bands. Conversely, O-mode photons become more dominant in harder bands. The number ratio between E-mode and O-mode photons, $N_E/N_O$, is higher for $\Theta=0$ than for $\Theta=\pi/2$, as a result of the suppression of the E-mode photon opacity in strong magnetic fields near the magnetic pole.

(ii.)  The polarization degree is zero for $\Theta=0$ and reaches a maximum value  for $\Theta=\pi/2$. The maximum polarization degree from the trapped fireball is lower than that ($\sim 100\%$) from a hot spot on the pole of the magnetar surface \citep{ade06,ade09} or in the quiescent magnetar magnetosphere \citep{fer11,tav14}. This is mostly due to the extended thermal emission from a trapped fireball and the wider collected band, e.g. $1-30\,\unit{keV}$ or $30-100\,\unit{keV}$ considered in our work.

(iii.) During the photon propagation in the magnetic vacuum, due to the adiabatic condition ($r\ll r_p$), the electric vector direction of the electromagnetic wave continues to change according to the direction of the local magnetic field.  Because the polarization-limit radius $r_p$ is much larger than that of the trapped fireball radius, the electric vectors of all the photons with a certain energy emitted from the photosphere rotate a same angle $\sim\phi(r_p)$ as they escape the magetosphere.

(iv.) The polarization degree at $\pi/2$ can be as large as $\sim 30\%$ in 1-30 keV, and is $\sim 10\%$ in 30-100 keV. Since giant flares typically have large fluxes, such a polarization degree would be detectable future X-ray/$\gamma$-ray polarimeters such as POLAR and LEAP, and may have been detected by the existing X-ray detectors such as RHESSI. We encourage a re-analysis of the possible polarization signal of SGR 1806-20 giant flare pulsating tail in the RHESSI data \citep{bog07}.

C. Implications for detecting the predicted polarization signal:

Our calculations are based on the magnetar conjecture and robust QED physics. If the future observations confirm the polarization signal at the predicted level, it would suggest the following:

(i.) The magnetar hypothesis is correct. Even if the magnetar model is the leading model to interpret SGRs, some other scenarios have been proposed to interpret SGRs without introducing strong magnetic fields. For example, in the accretion models \citep{alp01}, the strength of the magnetic field is weak, i.e. $B\simeq10^{11}\,\unit{G}$. According to Eq.(\ref{eq:QED-radius}), the QED polarization limiting radius would be $r_p\simeq 15R_s$, which is not much larger than the scale of the trapped fireball. This would give rise to a much smaller polarization degree.

(ii.) QED is right: A high polarization degree suggests that vacuum polarization indeed makes a corrected contribution to the dielectric tensor and the magnetic permeability tensor, and that the QED vacuum is birefringent. Otherwise the polarization degree would be less than $\sim 10\%$ even for $\Theta=\pi/2$.

(iii.) A polarization measurement would reveal the geometry of the system.
Since $N_E/N_O$ is of the order of unity, the polarization degree is mainly determined by the magnetic field line directions. Therefore, a relatively large polarization degree corresponds to an ordered magnetic field configuration, and a relatively large angle between the magnetic pole and line of sight.

(iv.) The idea of the pair-plasma trapped fireball is right: Different from the electron-ion plasma, the pair plasma has no vacuum resonance point and ion cyclotron absorption. Both E-mode and O-mode photons are entirely linearly polarized. The properties of pair opacity are also very different from that of electron-ion plasma. Therefore, an observed polarization degree at the predicted level would prove that the trapped fireball is indeed a photon-pair plasma.

In our Monte Carlo simulations, we have assumed an aligned rotator (the rotation axis and magnetic axis overlap), so that the $\Theta$ angle is constant for a particular observer. In reality, the two axes are mis-aligned. The slow rotation of the magnetars would not affect the $\Theta$-dependent polarization degrees calculated in this paper. However, a certain observer would view the magnetosphere in a range of $\Theta$ values. The average polarization degree would be between the minimum and maximum $\Pi$ values defined the maximum and minimum $\Theta$, which would usually be still $\sim 10\%$ in the 1-30 keV band (or a few percent in 30-100 keV), unless the two axes are nearly aligned and the line of sight is also very close to the pole.

We thank the anonymous referee for valuable and detailed suggestions that have allowed us to improve this manuscript significantly.
This work is partially supported by National Basic Research Program (973 Program) of China under Grant No. 2014CB845800. YPY is supported by China Scholarship Program to conduct research at UNLV.

\appendix

\section{Free electron gas in strong magnetic fields}

In order to obtain the structure of a trapped fireball, we need to know the relation between the pair energy density and temperature. Here, we summarize the basic properties of a free electron gas in strong magnetic fields. The results are consistent with \citet{tho95}.

Consider an elementary cell in a phase space with a volume
\begin{eqnarray}
\Delta x\Delta y\Delta z\Delta p_x\Delta p_y\Delta p_z=h^3,
\end{eqnarray}
where $(x,y,z)$ and $(p_x,p_y,p_z)$ are the position and momentum vectors of the electrons.
The electron distribution function as a function of energy $E$ is given by
\begin{eqnarray}
f=\frac{g}{e^{(E-\mu)/k_BT}+1},
\end{eqnarray}
where $g$ is the degeneracy number and $\mu$ is the chemical potential.
For 3-D free electrons, the number density is given by
\begin{eqnarray}
n_e&=&\int_0^\infty \frac{g}{e^{(E-\mu)/k_BT}+1}\frac{4\pi p^2}{h^3}dp.
\end{eqnarray}
In strong magnetic fields, the electron distribution is in 1-D (due to the strong Landau confinement), which is given by \citep{har06}
\begin{eqnarray}
n_e=\frac{1}{(2\pi r_c)^2\hbar}\sum_{n=0}^\infty g_n\int_{-\infty}^\infty \frac{1}{e^{(E_n-\mu)/k_BT}+1} dp_z,
\end{eqnarray}
where $r_c=(\hbar c/eB)^{1/2}$ is the cyclotron radius, and $g_n$ is the degeneracy at energy level $n$ ($g(n=0)=1$ and $g(n\geq 1)=2$). The Landau level in a strong magnetic field is given by
\begin{eqnarray}
E_n=\left[m_e^2c^4\left(1+2n\frac{B}{B_Q}\right)+p_z^2c^2\right]^{1/2}.
\end{eqnarray}
The first Landau level is
\begin{eqnarray}
E(n=1)\simeq\left\{
\begin{array}{lll}
m_ec^2(B/B_Q), &\,\,& (B\ll B_Q),\\
m_ec^2(2B/B_Q)^{1/2}, &\,\,& (B\gg B_Q).\\
\end{array}
\right.
\end{eqnarray}

i) If $E(n=1)\ll k_BT$, electrons are in the non-magnetic limit, and the higher Landau levels are occupied. Thus one should consider a 3-D Fermi distribution.

a. For $k_BT\ll m_ec^2$ ($B\ll B_Q$) and $\mu\ll m_ec^2$, one has
\begin{eqnarray} 
U_{e^\pm}=m_ec^2n_{e^\pm}&\simeq&2m_ec^2\frac{4\pi g}{h^3}e^{(\mu-mc^2)/k_BT}\int_0^\infty e^{-p^2/2m_ek_BT}p^2dp\nonumber\\
&\simeq&m_ec^2\frac{2^{1/2}}{\pi^{3/2}}\frac{(m_ec)^3}{\hbar^3}\left(\frac{k_BT}{mc^2}\right)^{3/2}e^{-m_ec^2/k_BT};
\end{eqnarray}

b. For $k_BT\gg m_ec^2$ and $\mu\ll pc$, one has
\begin{eqnarray}
U_{e^\pm}&\simeq&2\frac{4\pi g}{h^3}\int_0^\infty pc \frac{1}{e^{pc/k_BT}+1}p^2dp\nonumber\\
&=&\frac{14\pi^5}{15}\frac{(k_BT)^4}{(hc)^3}=\frac{7}{4}U_\gamma\label{a8}.
\end{eqnarray}

ii) If $E(n=1)\gg k_BT$, only the ground Landau level is occupied by electrons. One should consider a 1-D magnetized Fermi distribution.

c. For $k_BT\ll m_ec^2$ and $\mu\ll m_ec^2$, one has
\begin{eqnarray}
U_{e^\pm}=m_ec^2n_{e^\pm}&=&2m_ec^2\frac{1}{(2\pi r_c)^2\hbar}\sum_{n=0}^\infty g_n\int_{-\infty}^\infty \frac{1}{e^{(E_n-\mu)/k_BT}+1} dp_z\nonumber\\
&\simeq&2m_ec^2\frac{1}{(2\pi)^2\hbar}\frac{eB}{\hbar c}e^{(\mu-m_ec^2)/{k_BT}}\int_{-\infty}^\infty e^{-p_z^2/2m_ek_BT}dp_z\nonumber\\
&\simeq&m_ec^2\frac{(m_ec)^3}{\hbar^3(2\pi^3)^{1/2}}\frac{B}{B_Q}\left(\frac{k_BT}{m_ec^2}\right)^{1/2}e^{-m_ec^2/k_BT}; \label{a9}
\end{eqnarray}

d. For $k_BT\gg m_ec^2$ ($B\gg B_Q$) and $\mu\ll p_zc$, one has
\begin{eqnarray}
U_{e^\pm}&=&2\frac{1}{(2\pi r_c)^2\hbar}\sum_{n=0}^\infty g_n\int_{-\infty}^\infty p_zc \frac{1}{e^{(p_zc-\mu)/k_BT}+1} dp_z\nonumber\\
&\simeq&2\frac{1}{(2\pi)^2\hbar}\frac{eB}{\hbar c}\frac{(k_BT)^2}{c}2\int_0^\infty \frac{x}{e^x+1} dx\nonumber\\
&=&\frac{1}{12}\frac{B}{B_Q}\frac{m_e^2c}{\hbar^3}(k_BT)^2.
\end{eqnarray}

\clearpage

\begin{deluxetable}{@{\extracolsep{\fill} } ccccc}
\tabletypesize{\scriptsize} \tablecaption{The polarization degree of the trapped fireball} 
\tablewidth{0.8\textwidth}
\tablehead{\colhead{$\Theta^\mathrm{a}$} & \colhead{$(N_E/N_O)_S^\mathrm{b}$} & \colhead{$\Pi_S^\mathrm{c}$} &
\colhead{$(N_E/N_O)_H^\mathrm{d}$} & \colhead{$\Pi_H^\mathrm{e}$}} 
\startdata
0 		& 2.78 & 0.5\% & 1.06 & 0.9\%\\

$\pi/4$ & 2.33 & 12.7\% & 0.84 & 2.7\%\\

$\pi/2$ & 1.98 & 27.9\% & 0.80 & 10.0\%\\
\enddata
\tablenotetext{a}{$\Theta$ is the angle between the magnetic axis and the line of sight.}
\tablenotetext{b}{$(N_E/N_O)_S$ is the rate of the number of E-mode photons and O-mode photons at $1-30\,\unit{keV}$.}
\tablenotetext{c}{$\Pi_S$ is the polarization degree at $1-30\,\unit{keV}$.}
\tablenotetext{d}{$(N_E/N_O)_H$ is the rate of the number of E-mode photons and O-mode photons at $30-100\,\unit{keV}$.}
\tablenotetext{e}{$\Pi_H$ is the polarization degree at $30-100\,\unit{keV}$.}
\tablenotetext{\ast}{Note that E mode and O mode are defined by the direction of the magnetic field at $r=10R_s$ where they are under adiabatic condition.}
\label{tab1} 
\end{deluxetable}

\end{document}